\documentclass[a4paper,11pt]{article}
\usepackage[scale={0.8,0.9},centering,includeheadfoot]{geometry}
\usepackage{amssymb,amsmath, graphicx,epsfig,amscd,array,color}
\usepackage{amsthm}
\usepackage[round]{natbib}
\usepackage{lineno}
\usepackage{bbm}
\usepackage{authblk}
\usepackage{algorithm2e}
\newtheorem{theorem}{Theorem}

\newtheorem{proposition}{Proposition}

 \newcommand{\norm}[1]{\lVert#1\rVert}
 \def\mm#1{\ensuremath{\boldsymbol{#1}}} %

  \DeclareMathOperator{\tr}{tr} %

\begin{document}

\title{Scalable iterative methods for sampling from massive Gaussian random vectors}
\author[1]{Daniel P. Simpson\footnote{Corresponding author. Email: \texttt{daniel.simpson@math.ntnu.no}}}
\author[2]{Ian W. Turner}
\author[3]{Christopher M. Strickland}
\author[2]{Anthony N. Pettitt }

\affil[1]{Department of Mathematical Sciences, Norwegian University of Science and Technology, N-7491 Trondheim, Norway}
\affil[2]{Mathematical Sciences School, Queensland University of Technology, Brisbane, Australia}
\affil[3]{School of Economics, Australian Business School, University of New South Wales, Sydney, Australia}

\maketitle

\begin{abstract}
Sampling from Gaussian Markov random fields (GMRFs), that is multivariate Gaussian random vectors that are parameterised by the inverse of their covariance matrix, is a fundamental problem in computational statistics.  In this paper, we show how we can exploit arbitrarily accurate approximations to a GMRF to speed up Krylov subspace sampling methods.  We also show that these methods can be used when computing the normalising constant of a large multivariate Gaussian distribution, which is needed for both any likelihood-based inference method. The method we derive is also applicable to other structured Gaussian random vectors and, in particular, we show that when the precision matrix is a perturbation of a (block) circulant matrix, it is still possible to derive $\mathcal{O}(n\log n)$ sampling schemes. 
\end{abstract}

\paragraph{Keywords:} Gaussian Markov random field; Lanczos algorithm; Krylov subspace; Hutchinson estimator;  Markov chain Monte Carlo; Super-geometric convergence; Log-Gaussian Cox process.

\section{Introduction}

Sampling from large multivariate Gaussian random vectors lies at the heart of any number of tools for performing
Bayesian inference. In particular, it is typically a fundamental operation in a number of popular Markov chain Monte Carlo (MCMC)
methods, such as random walk Metropolis, Metropolis adjusted Langevin, and Hamiltonian Monte Carlo algorithms.  When
the dimension of the target distribution is large, sampling becomes a computational bottleneck and it is no longer
possible, in a reasonable time frame, to use standard methods to construct samples.  In this paper, we propose a new
method for performing inference on models with large Gaussian components that remains feasible even when the model
under consideration is massive.  These methods use a controlled amount of memory and can be used to compute a sample up to arbitrary precision.

In order to obtain solutions to many high-dimensional problems within a reasonable computational budget, it is necessary to
introduce additional structure to both the model and the inferential scheme. For example, in order to make models in
spatial statistics computationally feasible, one is forced to make assumptions about the independence structure
\citep{art456, kaufman2008covariance}, the conditional independence structure \citep{book80,Lindgren2011}, or enforce
some sort of low dimensional structure \citep{art473,art445,art444}.  These assumptions, which attempt to balance
computational realities with modelling flexibility, allow statisticians to fit models to relatively large spatial data sets, and to compute
reasonably high-resolution spatial prediction surfaces.    However, it is not uncommon to come across data sets for which
these methods are not sufficient, especially when looking at space-time satellite data \citep{Strickland2010} or three dimensional problems \citep{Aune2011}.  The computational bottleneck comes in the matrix operations required to evaluate a Gaussian likelihood, compute a proposal for the Gaussian component, and for computing spatial estimates. 

Given the importance of the problem, there are a large number of methods for sampling from Gaussian random vectors $\mm{x} \sim N(\mm{0},\mm{Q}^{-1})$.   Typically, they revolve around computing a factorisation of the form $\mm{Q} = \mm{LL}^T$ and noting that $\mm{L}^T\mm{x} \sim N(\mm{0},\mm{I})$.  The standard choice is to use the Cholesky factorisation of $\mm{Q}$, in which case $\mm{L}$ is lower triangular \citep{book80}.  Another option is to chose $\mm{L}$ to be the matrix square root of $\mm{Q}$, however this will only be feasible when $\mm{Q}$ can be cheaply diagonalised.  In particular, this is the case in the important situation where $\mm{Q}$ is circulant.  The problem with this approach is that, as the dimension of $\mm{x}$ increases, the cost of the matrix factorisation increases sharply. In the most general case, the number of floating point operations grows as $\mathcal{O}(n^3)$, while the memory costs grow like $\mathcal{O}(n^2)$, where $n$ is the dimension of $\mm{x}$.     On modern computers, the quadratic growth in memory will render large problems completely impossible.

In  order to avoid this problem, we will focus  on a class of iterative methods, known as Krylov subspace methods, initially introduced by \citet{Simpson2007} and further extended and applied by \citet{Strickland2010} and \citet{Aune2011,Aune2012}.   These methods do not require a direct factorisation (or even storage) of the matrix, but instead use modern numerical linear algebra techniques to compute $\mm{x}_m \approx \mm{Q}^{-1/2}\mm{z}$, where $\mm{z}$ is a vector of i.i.d. standard normal random variables and the parameter $m$ can be chosen adaptively to control the error in the approximation.  These methods only require $\mathcal{O}(n)$ storage, which means that they will remain feasible for far larger problems than direct methods.   Unfortunately, the number of steps that the algorithm requires to reach a prescribed error level grows polynomially in $n$, which makes them very time consuming in practice.  In this article, we significantly extend the methods developed by \citet{Simpson2007} by developing tools that slow and, in some cases, completely remove this dimension-dependent cost increase.  This allows us to finally construct matrix-free, dimension independent samplers for structured Gaussian random vectors.  

In this paper, we will show how the structure of the Gaussian random vector can be used to make efficient, dimension-independent samplers.  The paper proceeds as follows.  In Section \ref{sec:structure}, we briefly review the types of structure commonly found in Gaussian random vectors, while in Section \ref{sec:LGCP} we discuss efficient methods to conditionally sample from a log-Gaussian Cox process.  This example will run throughout the paper and be used to demonstrate various algorithms developed in this paper.  The basic Krylov sampler, as introduced in \citet{Simpson2007}, is recounted in Section \ref{sec:krylov}.  It is shown that, even though it converges to the true sample faster than geometrically, this method does not scale well with the size of the problem.  In Section \ref{sec:preconditioning}, we develop a new method that improves the scaling of the method.  In particular, though a careful re-parameterisation we can make these methods scale perfectly with dimension.  We provide methods for constructing optimal, super-geometric, dimension independent, arbitrarily accurate $\mathcal{O}(n\log n)$ samplers in the case where $\mm{Q}$ is a bounded perturbation of a block circulant or block Toeplitz matrix. Guidance is also provided for building good re-parameterisations of other Gaussian random vectors, however the optimal choice is still an open research question.   Although building a good  sampler is an important problem, in applications it usually necessary to also be able to compute the log-density of the multivariate Gaussian random vector.  The computational bottleneck here is the computation of the log-determinant of a massive matrix and in Section \ref{sec:determinant} we extend the seminal work of \citet{Hutchinson1990} and \citet{bai1996some}, as well as a some more recent work by \citet{Aune2012}, to show that we can use similar re-parameterisations to construct variance-reduced Monte Carlo estimator that have dimension-independent relative error.  As the estimates of the log-likelihood come from a Monte Carlo scheme, it will never be computed to high accuracy and, in Section \ref{sec:exactness}, we discuss the effect of this inexactness on inference.   Finally, in Section \ref{sec:conclusion}, we summarise the work presented in this paper and discuss some future directions.

\section{Structured Gaussian random vectors} \label{sec:structure}

In order to motivate the methods considered in this paper, it is useful to take a closer look at the types of computationally efficient modelling structures that frequently occur.   There are three common ways that $\mm{Q}$ can be structured in order to simplify computations with multivariate Gaussian random vectors.  If the precision matrix $\mm{Q}$ is \emph{sparse}, then this corresponds to a Markovian dependence structure between components of
the random vector and such multivariate Gaussians are known as Gaussian Markov random fields (GMRFs) \citep{book80}.  In
this case, powerful methods from sparse linear algebra can be used to speed up computations and, when the dependence is
spatial, the cost is commonly $\mathcal{O}(n^{3/2})$ \citep{art192}.    The second common structure for multivariate
Gaussians occurs when the  precision matrix is circulant or block circulant. These models are classically used when
considering spatial models over large, regular lattices \citep{book80,moelleral:98}.  As block circulant matrices can be
diagonalised using fast Fourier transforms, all calculations with these multivariate Gaussians can be performed in
$\mathcal{O}(n\log n)$ operations.   The third common structure for multivariate Gaussians occurs when using Gaussian
random fields modelled on finite dimensional stochastic processes \citep{art445,art444}, in which case the covariance
matrix is typically a sparse matrix added to a low-rank matrix.  In this case, the Sherman-Morrison-Woodbury formula can
be used in the computations and the cost is usually $\mathcal{O}(nr^3)$, where $r$ is the rank of the perturbation. 

The three classes of models discussed in the previous paragraph share a common characteristic:  it is cheap to compute the
matrix-vector product $\mm{Qv}$ for any vector $\mm{v} \in \mathbb{R}^n$.  In fact, it is often easy to write a routine
that computes the matrix-vector product without ever forming or storing the matrix $\mm{Q}$.  In particular, the
matrix-vector products for the three models cost, respectively, $\mathcal{O}(n)$, $\mathcal{O}(n\log n)$, and
$\mathcal{O}(nr)$ operations.  Furthermore, if the precision matrix of a model is, say, a circulant matrix added to a
sparse matrix, it is still possible to form cheap matrix-vector products even though the model itself no longer has a
special structure that classical algorithms can take advantage of.

\section{Motivating example: A good MCMC sampler for log-Gaussian Cox processes} \label{sec:LGCP}

The methods described in this paper are designed to solve high-dimensional problems.  While these problems arise in a
number of interesting contexts, see, for example, the literature on animal breeding \citep{Gorjanc2010Graphical}, for
simplicity we focus on problems in spatial statistics.  In particular, we focus on inference for log-Gaussian Cox
processes (LGCPs).  This problem has all of the structure of the general type of problem that our methods will handle
well, while having enough analytical structure to get results that can be used to build intuition in the more general
case.  

Conditional sampling from LGCPs, that is a Poisson point process for which the log intensity surface is modelled through
a Gaussian random field, is a challenging problem for MCMC methods.  This is a very high (actually infinite) dimensional
sampling problem and, as such, it is difficult to design an MCMC scheme that efficiently explores the posterior.
Given an observed point pattern $Y$, the likelihood for a LGCP can be written in hierarchical form as  \begin{align*}
\pi({Y} | x(\cdot))&= \exp\left(|\Omega| - \int_{\Omega} \exp(x(s)) \,ds\right)\prod_{s_i \in Y} \exp(x(s_i)) \\
		x(\cdot) & \sim GRF(\mu(\cdot), c(\cdot,\cdot)), \end{align*} where $ GRF(\mu(\cdot), c(\cdot,\cdot))$ is a
Gaussian random field with mean function $\mu(\cdot)$ and covariance function $c(\cdot,\cdot)$.  The integral in
the likelihood cannot be computed analytically and, therefore, this likelihood is ``doubly intractable''.
Although there are a number of methods for resolving this intractability \citep[see][for a
survey]{Girolami2013Playing}, in this paper we will follow standard practice \citep{art245} and approximate the
likelihood.  A simple way to approximate the likelihood is to  discretise it over a computational lattice that
covers the observation window and approximate the model with the latent Gaussian model
\begin{subequations}\label{LGCP} \begin{align} y_{ij} | x_{ij} & \sim Po\left(e^{x_{ij}}\right) \\ \mm{x} &\sim
N(\mm{0}, \mm{Q}^{-1}), \end{align} \end{subequations} where  $y_{ij}$ is the number of points in the $(i,j)$th
cell and $\mm{x} \in \mathbb{R}^n$ is a stationary random field over the lattice that, for convenience, we will
take to be defined on a torus with block circulant precision matrix $\mm{Q}$.  It can be shown that this
approximation converges as the lattice is refined \citep{waagepetersen2004}.

The interesting thing about inferring LGCPs, in the context of this paper, is that the lattice structure is artificially
imposed and should in practice be taken to be as fine as possible.  Therefore, we are interested in developing methods
that continue to work well when the dimension of the latent field is enormous.  To see why this is a challenge, 
consider the structure of the prior distribution placed on $\mm{x}$.   Due to the extremely informative nature of infinite dimensional priors and the relative lack of information present in a point pattern, it is expected that the posterior will be largely determined by the (discretised) Gaussian random field prior that has been placed  on $\mm{x}$.   In fact, the lack of an infinite dimensional analogue of a Lebesgue
measure means that the prior distribution $\pi(\mm{x})$ will become singular as its dimension increases.  Thanks to the
assumption that the driving process $x(\cdot)$ is stationary on a torus, we can actually track just how singular
$\mm{Q}$ becomes.  It can be shown, without too much effort, that the condition number of $\mm{Q}$, that is the ratio of its largest and smallest eigenvalues grows like  $\mathcal{O} (h^{-(d+2\nu)})$, where $h$ is the size of the lattice and $d$ is the dimension of the problem (hereafter taken to be equal to two), and $\nu$ is the mean square smoothness of $x(\cdot)$.  This suggests that, as the lattice is refined, the problem becomes harder in floating point arithmetic \citep{higham1996accuracy} and that Gibbs samplers for sampling from $\mm{x}$ will converge slower \citep{art96,Parker2012Convergence} as the dimension increases.  We note in passing that we can get similar growth rates from non-lattice approximations to $x(\cdot)$ \citep{Lindgren2011} and that these can also be used to approximate log-Gaussian Cox processes \citep{SimpsonLGCP}

There has been an massive amount of work done on efficient MCMC methods for log-Gaussian Cox processes and the most commonly used method appears to be the preconditioned Metropolis-adjusted Langevin algorithm (MALA), which has the proposal
\begin{equation} \label{MALA}
\mm{x}^* | \mm{x} \sim N\left(\mm{x} + \frac{\delta^2}{2} \mm{Q}^{-1} \nabla_{\mm{x}} \log\left( \pi(\mm{y}|\mm{x})\right), \delta^2\mm{Q}^{-1}\right).
\end{equation}
The main advantage of this sampler is that, due to the block circulant structure of $\mm{Q}$, a proposal can be drawn using only $\mathcal{O}(n\log n)$ floating point operations and $\mathcal{O}(n)$ storage.  It is, however, well accepted in the MCMC literature that when sampling from latent Gaussian models, superior samplers can be constructed by exploiting likelihood information \citep{art192,art412,girolami2011riemann}.  To this end, we look at the simplified manifold MALA (sMMALA) scheme of \citet{girolami2011riemann}, which has the proposal 
\begin{equation} \label{sMMALA}
\mm{x}^* | \mm{x} \sim N\left(\mm{x} + \frac{\delta^2}{2} \left(\mm{Q} + \mm{H}\right)^{-1} \nabla_{\mm{x}} \log\left( \pi(\mm{y}|\mm{x})\right), \delta^2(\mm{Q} + \mm{H})^{-1}\right),
\end{equation}
where $\mm{H}$ is the Fisher information matrix of $\mm{y}|\mm{x}$, which is, in this case, diagonal.  While this sampler performs better than the vanilla MALA \citep{girolami2011riemann}, the precision matrix no longer has block circulant structure and therefore requires $\mathcal{O}(n^3)$ floating point operations and $\mathcal{O}(n^2)$ storage to generate a proposal.  Clearly this is not a feasible sampler for large lattices.

In the following sections we will show that if we carefully construct an iterative sampler, we leverage the remaining computational structure to  generate a proposal from \eqref{sMMALA} using only $\mathcal{O}(n\log n)$ floating point operations and $\mathcal{O}(n)$ storage, albeit with larger suppressed constants. Therefore, it is possible to use the superior proposal scheme \eqref{sMMALA} without sacrificing the exemplary computational properties of the inferior proposal \eqref{MALA}.

\section{Krylov subspace methods for sampling from Gaussian random vectors} \label{sec:krylov}

\begin{figure}
\centering
\includegraphics[width=0.7\textwidth]{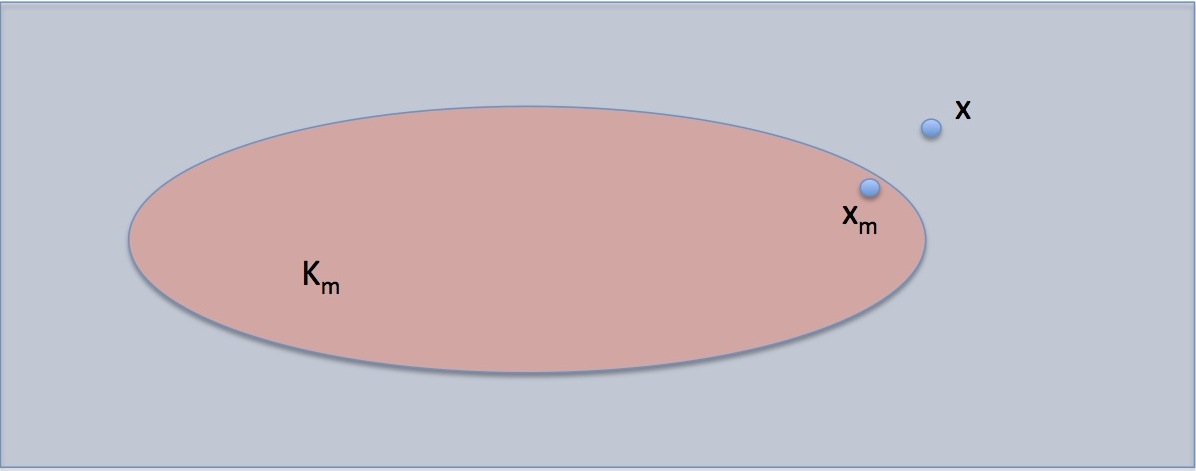}
\caption{This figure shows a schematic representation of the Krylov sampler.  Given a target sample $\mm{x} = \mm{Q}^{-1/2}\mm{z}$, the sampler constructs a sequence of subspaces $\mathcal{K}_m(\mm{Q},\mm{z})$ and computes a near-optimal approximation to $\mm{x}$, denoted $\mm{x}_m$ within this subspace.  This estimate depends non-linearly on $\mm{z}$.  \label{fig:krylov_schematic}}
\end{figure}

When it is possible to inexpensively compute the matrix-vector product $\mm{Qv}$ for arbitrary vectors $\mm{v}$, a Krylov subspace method can be constructed for sampling from $\mm{x} \sim N(\mm{0},\mm{Q}^{-1})$ \citep{Simpson2007,Simpsonthesis,Strickland2010,Ilic2009,Aune2011}.  This sampler is based off the observation that, if $\mm{z} \sim N(\mm{0},\mm{I})$ is a vector of independently and identically distributed normal variables, $\mm{x} = \mm{Q}^{-1/2}\mm{z}$ is a multivariate Gaussian with precision matrix $\mm{Q}$, where $\mm{Q}^{-1/2}$ denotes the inverse of the principle square root of $\mm{Q}$.   The method is then based on constructing a sequence of good approximations to $\mm{Q}^{-1/2}\mm{z}$ for a fixed realisation of $\mm{z}$.   This makes our method, which is illustrated graphically in Figure \ref{fig:krylov_schematic}, substantially different to the Gibbs sampler-based methods analysed by \cite{art96}, which produce Markov chains that converge geometrically in distribution to $N(\mm{0},\mm{Q}^{-1})$.  In contrast to this, we will see that due to the adaptive nature of our sampler it converges to the targeted sample faster than geometrically.

At its heart, the Krylov sampler, which is described in Algorithm  \ref{alor:lanczos}, is a dimension reduction technique, where the sampling problem is projected onto a low-dimensional space that is sequentially  constructed in such a way that it contains the main features of both the precision matrix $\mm{Q}$ and the noise vector $\mm{z}$.  This idea is the basis for both the ubiquitous conjugate gradient method for solving linear systems \citep{book43} and the partial least squares method in applied statistics \citep{wold1984collinearity}.  In fact, it can be shown that the convergence of the Krylov sampler mirrors the convergence of the conjugate gradient method as the following theorem, which is proved in a more general form in \citet{Ilic2009}, demonstrates.  This bound can also be extended in a fairly straightforward manner to finite precision arithmetic \citep{Simpsonthesis}.

\begin{theorem} \label{thm:convergence}
Let $\mm{x}_m$ be the sample produced in the $m$th step of the Krylov sampler and let $\mm{x} = \mm{Q}^{-1/2}\mm{z}$ be the true sample from $\mm{x}\sim N(\mm{0},\mm{Q}^{-1})$.  If $\mm{r}_m$ is the residual  at the $m$th iteration of the conjugate gradient method for solving $\mm{Qy} = \mm{z}$, then 
\begin{equation} \label{krylov_bound}
\norm{\mm{x} -\mm{ x}_m} \leq \lambda_{min}^{-1/2}\norm{\mm{r}_m},
\end{equation} 
where $\lambda_{min}$ is the smallest eigenvalue of $\mm{Q}$. Furthermore, the following \emph{a priori} bound holds:
\begin{equation} \label{apriori}
\norm{\mm{x} - \mm{x}_m} \leq 2\lambda_{min}^{-1/2}\sqrt{\kappa}\left(\frac{\sqrt{\kappa} -1}{\sqrt{\kappa} + 1}\right)^m \norm{z},
\end{equation} and $\kappa = \lambda_{max}/\lambda_{min}$ is the condition number of $\mm{Q}$.
\end{theorem}

\begin{algorithm}[tbp]

\hrule\hrule\hrule\vspace{10pt}
\KwIn{The precision matrix $\mm{Q}$ and the subspace size $m$.}
\KwOut{$\mm{x}_m$, an approximate sample from $N(\mm{0}, \mm{Q}^{-1})$} 
\vspace{10pt}\hrule\vspace{10pt}
Sample $\mm{z} \sim N(\mm{0},\mm{I})$\;
Set $\mm{v}_1 =\mm{z}/\norm{\mm{z}}$\;
\For{$j=1:m$}{
	Set $\mm{q} = \mm{Qv}_j$\;
	\If{$j\neq 1$}{
		$\mm{q} = \mm{q} - \beta_{j-1}\mm{v}_{j-1}$\;
	}%
	$\alpha_j = \mm{v}_j^T\mm{q}$\;
	$\mm{q} = \mm{q} -\alpha_j\mm{v}_j$\; 
	$\beta_j = \norm{\mm{q}}_2$\;
	$\mm{v}_{j+1} = \mm{q}/\beta_j$ \;

}%
Form $\mm{V}_m = \left[ \mm{v}_1, \mm{v}_2,\ldots,\mm{v}_m\right]$ and the symmetric tridiagonal matrix $\mm{T}_m$ with diagonal entries $\alpha_j$ and sub/super-diagonal entries $\beta_j$\;
Set $\mm{x}_m =\norm{\mm{z}} \mm{V}_m \mm{T}_m^{-1/2} \mm{e}_1$, where $\mm{e}_1 = (1, 0,\ldots, 0)^T \in \mathbb{R}^m$.

\vspace{10pt}\hrule\hrule\hrule\vspace{10pt}
\caption{The Lanczos algorithm for approximately sampling from $N(\mm{0} ,\mm{Q}^{-1})$.  In practice, the square root of the $m \times m$ tridiagonal matrix in the last step of the algorithm can be replaced with its rational approximation, which reduces the complexity of the final step on the sampler from $\mathcal{O}(nm + m^3)$ to $\mathcal{O}(nm)$.  \label{alor:lanczos}}
\end{algorithm}

 The  \emph{a priori} bound \eqref{apriori} implies that the number of iterations required to  prescribed error level grows log-linearly in $\sqrt{\kappa}$.  While this bound gives useful information about the qualitative convergence of the Krylov sampler, it is famously loose. Practically speaking, the valuable result in the theorem is the \emph{a posteriori} bound \eqref{krylov_bound}, which shows that the Krylov sampler behaves in the same manner as the conjugate gradient method.  Not only is this bound sufficiently tight that it can be used to evaluate the error in the Krylov sampler, but it also means that we can take  the insight gathered from sixty years of practical experience with the conjugate gradient method and transfer it directly to the Krylov sampler.  In particular, we know that the error will decrease ``superlinearly'', that is the error will behave like $\norm{\mm{x} - \mm{x}_m}  = \text{\cal{o}}(\rho^m)$ as $m \rightarrow \infty$ for any $\rho \in (0,1)$ \citep{simoncini2005occurrence}.  This means that the Lanczos sampler will converge faster than \emph{any} geometrically ergodic MCMC scheme for sampling from a multivariate Gaussian!  In practice, expectations must be tempered against the challenges of floating point arithmetic, however experience suggests that the error still displays superlinear behaviour up to the point at which the error stops decreasing.

\subsection{Running example: The behaviour of the Krylov sampler for sMMALA}

\begin{figure}
\centering
\includegraphics[width=0.7\textwidth]{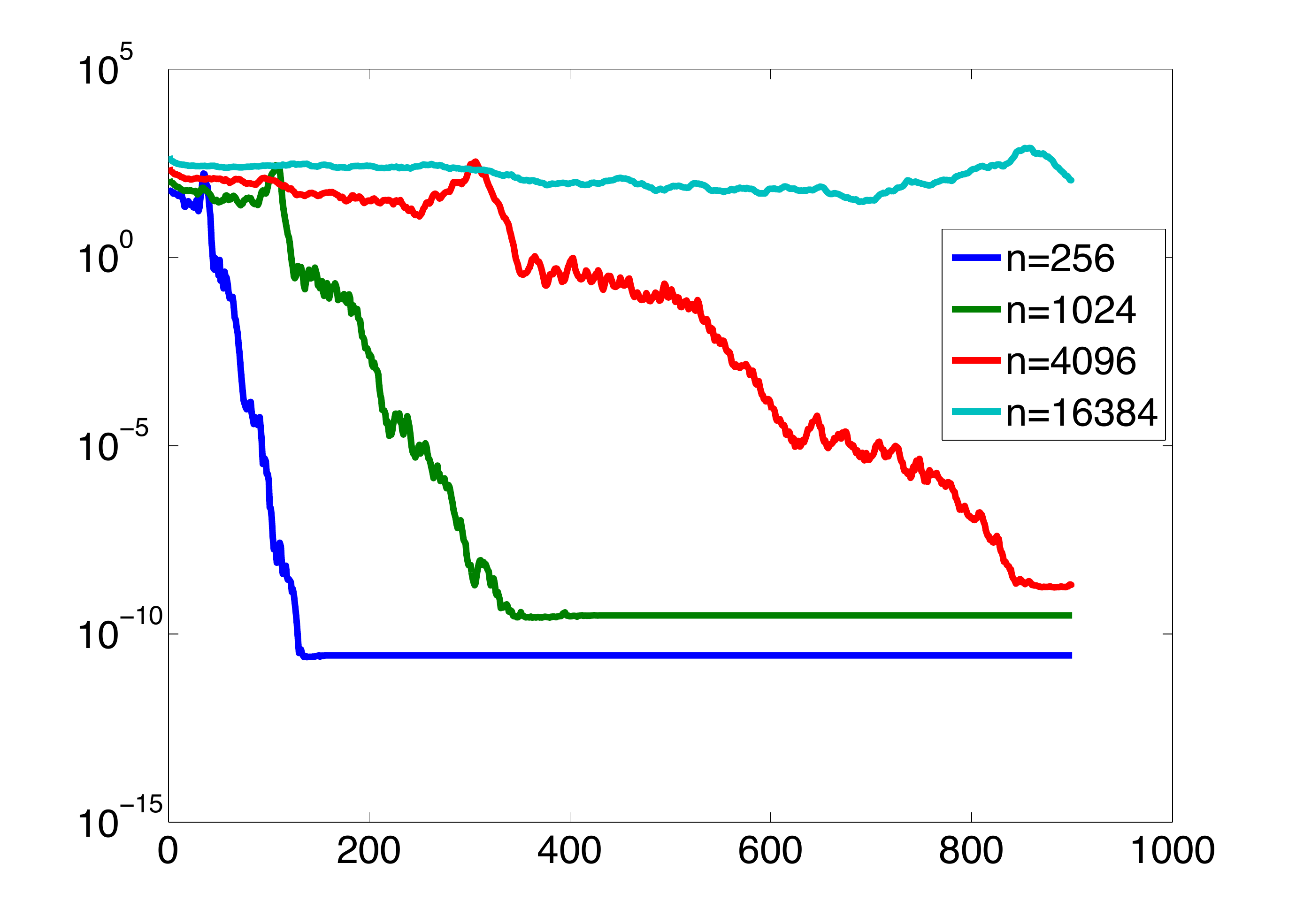}
\caption{This figure shows the convergence of Krylov sampling applied directly to the proposal \eqref{sMMALA}.  The x-axis shows the subspace size, while the y-axis shows an upper bound on the error (computed with \eqref{krylov_bound}). \label{fig:unpreconditioned}}
\end{figure}

The basic problem with the Krylov sampler, as suggested in Theorem \ref{thm:convergence}, is that the size of the Krylov subspace required to capture a good approximate sample increases with the condition number of the precision matrix.  In the case of our running example, as the lattice becomes denser we expect the performance of the Krylov sampler to degrade.  Figure \ref{fig:unpreconditioned}, which shows the decay of the upper bound \eqref{krylov_bound} as the size of the lattice increases, confirms that this is indeed the case.  For the largest lattice ($1024 \times 1024$), a $900$ dimensional subspace is not enough to generate a good approximation.

We note that the upper bounds that are plotted are slightly misleading, in that the true error is known to be non-increasing and it is observed empirically that the point where the superlinear convergence begins (when the rate of decrease gets faster) occurs in the true error earlier than it does in the bound.  However, the bound in Theorem \ref{thm:convergence} is that tightest bound the we have available and, therefore, the only available way of assessing convergence in practice.

\section{Improving the efficiency of the Krylov sampler: A preconditioning approach} \label{sec:preconditioning}

Although the bound in Theorem \ref{thm:convergence} suggests that the Krylov sampler may require a large number of iterations to converge, there is still hope.  When solving linear systems, such as those required to compute the mean of the proposal \eqref{sMMALA}, the slow convergence of the conjugate gradient method can be circumvented by a \emph{preconditioning} method, in which the linear system $\mm{Qu} = \mm{z}$ is replaced with $\mm{Q}\mm{M}^{-1}\mm{u} = \mm{z}$. The choice of the precondition $\mm{M} \approx \mm{Q}$ is vital to the method and it is chosen so that it is easy to invert \citep{book43}.  For a well chosen preconditioner, the condition number of $\mm{M}^{-1}\mm{Q}$ will be close to 1, and the conjugate gradient method on the preconditioned system will then only require a few iterations for convergence.

Given that it is not possible to apply a general preconditioner built for a linear system to computing a matrix function,  the preconditioning operation for the Krylov sampler is more delicate.  For a linear system, the fundamental property of a practical preconditioner is that it is possible to compute $\mm{M}^{-1}\mm{b}$ quickly for any vector $\mm{b}$.  The corresponding fundamental property when preconditioning the Krylov sampling turns out, unsurprisingly, to be sampling efficiently from $N(\mm{0},\mm{M}^{-1})$. This is obviously a much more difficult problem and essentially limits the types of preconditioners available to those that can be factored.  

Given that we can find a matrix $\mm{M}$ such that we can sample efficiently from $N(\mm{0},\mm{M}^{-1})$, the following proposition, which can be verified using the properties of the multivariate normal distribution,  outlines the method of preconditioning the Krylov sampler.

\begin{proposition} \label{prop:preconditioning}
Let $\mm{Q}$ and $\mm{M}$ be symmetric positive definite matrices and let $\mm{Q} = \mm{L}\mm{L}^T$ and $\mm{M} = \mm{F}\mm{F}^T$ be given decompositions.  If $\mm{u} \sim N\left(\mm{0}, \left(\mm{F}^{-1}\mm{Q}\mm{F}^{-T}\right)^{-1}\right)$, then the solution to $\mm{F}^T\mm{x} = \mm{u}$ is a zero-mean Gaussian random vector with precision matrix $\mm{Q}$.
\end{proposition}

If the preconditioner is perfect, that is if $\mm{M} = \mm{Q}$, then $\mm{u}$ will be a vector of i.i.d. normals and the proposition collapses to the standard method for sampling from Gaussian random vectors \citep{book80}.  Therefore, the dependency of $\mm{u}$ is a measure of how well $\mm{M}$ captures the essential properties of $\mm{Q}$.  The key property of $\mm{u}$ is that the spectrum of $\mm{F}^{-1} \mm{Q}\mm{F}^{-T}$ will usually be much more clustered than the spectrum of $\mm{Q}$ and, therefore, the Krylov sampler will converge faster.  

While Proposition \ref{prop:preconditioning} shows that we can replace the original sampling problem by one that may be easier to solve, we only have the vague guidance of Theorem \ref{thm:convergence}, which suggests that we should make the condition number of $\mm{M}^{-1}\mm{Q}$ small, to help us choose $\mm{M}$.  In the remainder of this section, we present two specific choices of $\mm{M}$ that, in turn, show the best-case and the more common place behaviour of preconditioned samplers.

\subsection{Running example: Fast sampling from circulant--plus--sparse matrices}

Generating a proposal from \eqref{sMMALA} requires a method to compute $ \left(\mm{Q} + \mm{H}\right)^{-1} \nabla_{\mm{x}} \log\left( \pi(\mm{y}|\mm{x})\right)$ and to sample from $N\left(\mm{0},(\mm{Q} + \mm{H})^{-1}\right)$, where $\mm{Q}$ is block circulant and $\mm{H}$ is sparse.  In this section, we will show that one preconditioner can be used to solve both problems.  In particular, this preconditioned is \emph{optimal} in the sense the the condition number of the preconditioned matrix does not depend on $h$.  This means that  the number of steps required for the  Krylov sampler to generate a realisation of a random field up to a given accuracy depends only on the accuracy and not on the size of the mesh!  %

The following theorem shows that preconditioning with the prior is, in many cases, sufficient for optimal convergence.  
\begin{theorem} \label{thm:converge_LGCP}
Let $Y$ be a log-Gaussian Cox process driven by a Gaussian random field $x(s)$ defined on the flat torus $W = [0,1]^2$ with stationary covariance function $c(h)$.   Let $\mm{y}$ be a vector of  $k<\infty $ realisations the discretised LGCP on an $n_1 \times n_2$ lattice and let $\mm{x}$ be  the discretisation of $x(s)$ over the same lattice.  If $\mm{x}_m$ is the approximation to a preconditioned sample $\mm{x}$ from the sMMALA proposal with preconditioner $\mm{M}_n = \mm{Q}_n + \alpha_n \mm{I}$, then $$
\norm{\mm{u} - \mm{u}_m }\leq C \left(\frac{ \left(\int_W \exp(x(s))\,ds  - C_\alpha\right)}{m}\right)^{m},
$$
where $C$ is a constant independent of $n=n_1n_2$ and $C_\alpha = \lim_{n\rightarrow \infty} n\alpha_n$.  Hence if $C_\alpha$ is finite, then  the sMMALA proposal \eqref{sMMALA} can be generated to any accuracy in $\mathcal{O}(n \log n)$ iterations.
\end{theorem}

The proof, which is given in Appendix \ref{appendix:proof}, relies on the prior dominating the data in the sense that $\mm{Q}_n^{-1}\mm{H}_n$ is bounded.  This  is commonly the case in practical spatial analysis (there are no uninformative infinite dimensional priors!), however in the rare case where there is enough data to overcome the prior, the same result holds by choosing $\mm{M} = \mm{H}$ instead of $\mm{M} = \mm{Q}$.

While Theorem \ref{thm:converge_LGCP} shows that a sampler requires $\mathcal{O}(1)$ steps to reach a fixed accuracy, the bound in the proof is very loose and does not guarantee that the number of steps required is small enough to be of practical use.  In Table 1, however, we show that, for the running example, the required number of iterations is quite small.  In this case $\alpha_n=0$ was chosen, however in practice, we can tune this parameter if required.  From a practical point of view, 6 iterations of the Krylov sampler requires $24$ FFTs, in contrast to the $2$ required to solve a circulant linear system.

\begin{table}
\begin{center}
\begin{tabular}{|c|ccccccccc|}
\hline
$m$ ($m \times m$ grid) & $16$  & $32$&   $64$ & $128$  &$256$ & $512$ &$1024$& $2048$ &$4096$ \\
\hline Preconditioned & $6$ & $6$ & $6$ & $6$ & $6$ & $6$ & $6$ & $6$ & $6$ \\
 Unpreconditioned & $102$ & $286$  & $790$  & $2166$  & - & - & - & -& - \\
\hline
\end{tabular}
\end{center}
\caption{This table, which gives the number of iterations required for the error bound \eqref{krylov_bound} to be less than $10^{-8}$, demonstrates that the preconditioned sampler indeed requires  $\mathcal{O}(n\log n )$ operations to reach a fixed accuracy. \label{table:circulant}}

\end{table}

\subsection{Preconditioning general problems} \label{sec:general}
When $\mm{Q}$ is sparse, there are several generic (or, in the nomenclature of numerical linear algebra, ``algebraic'')  choices of preconditioner that are available.  Unlike the preconditioner considered in the previous section, algebraic preconditioners are constructed from information about the structure of the problem (sparsity pattern, block structure, \emph{etc}), rather than the sort of detailed analytic knowledge used to construct the optimal preconditioner in the previous section.

The most obvious candidate for a generic preconditioner is the incomplete Cholesky decomposition \citep{book43} of $\mm{Q}$, which computes an approximation to the true Cholesky decomposition using a lower amount of fill in.  It was shown by \citet{art364} \citep[c.f.][]{Hu2012} that incomplete Cholesky decompositions can be used to approximate a fixed Gaussian Markov random field.  It is, therefore, expected that the incomplete Cholesky will be a successful candidate for a preconditioner in Proposition \ref{prop:preconditioning}.   A similar class of preconditioner is the  factored sparse approximate inverses \citep{kharchenko2001robust}, which can be constructed in parallel in a columnwise manner.  It is also possible to build preconditioners based on symmetric sweeps of stationary iterative methods \citep{book43}, which have strong connections to block Gibbs samplers in the Gaussian setting \citep{Parker2012Convergence}.

\begin{figure}
\centering
\includegraphics[width=0.7\textwidth]{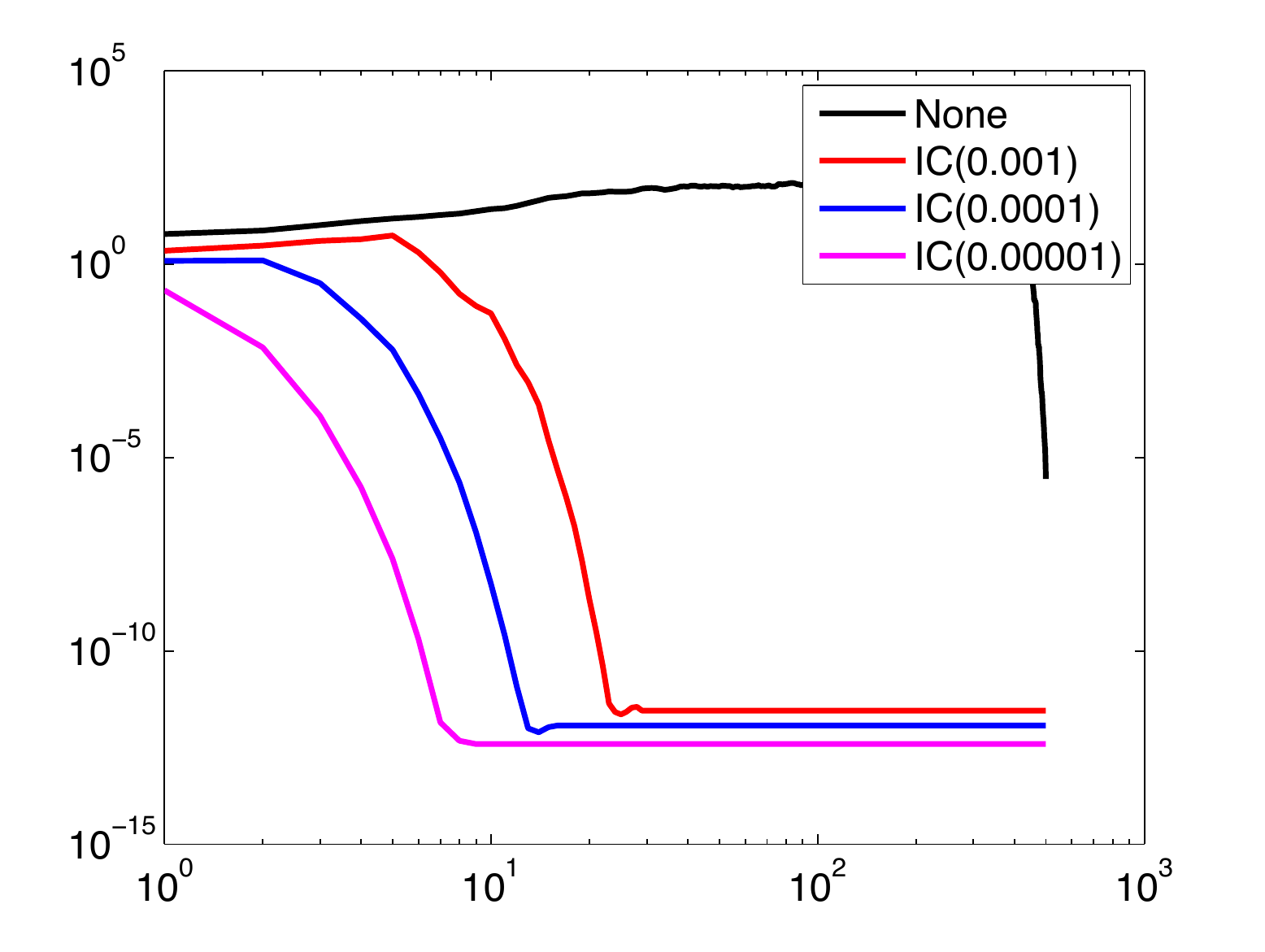}
\caption{A comparison of incomplete Cholesky preconditioners with several thresholds.  The x-axis shows the subspace size, while the y-axis shows an upper bound on the error (computed with \eqref{krylov_bound}). \label{fig:incomplete}}
\end{figure}

 In Figure \ref{fig:incomplete}, the convergence for the preconditioned Krylov sampler is shown for several variants of
 the incomplete Cholesky factorisation.  The test is performed on the square of the matrix constructed using the \textsc{Matlab}
 command \verb|Q = (31^2*gallery('poisson',30))^2|, which corresponds to a second order random walk on a $30 \times 30$
 lattice with a modification on the boundary to ensure that the distribution is proper \cite[see][for a
 definition]{book80}. In particular, we compare incomplete Cholesky factorisations with various  thresholding levels and it is clear that  this  simple re-parameterisation can greatly improve the performance of the Krylov sampler.

\subsection{Connection with non-centred parameterisation}
The preconditioners considered in this paper are closely linked to the concept of `centred' and `non-centred' parameteristations of statistical models \citep{papaspiliopoulos2007general, strickland2008parameterisation,yu2011center,FilipponeML13}.   The idea can be illustrated simply for  parameter-dependent latent Gaussian models
\begin{align*}
\mm{y} | \mm{x},\mm{\theta}  &\sim \pi(\mm{y} | \mm{x},\mm{\theta}) \\
\mm{x} | \mm{\theta} &\sim N(\mm{0}, \mm{Q}(\mm{\theta)}^{-1}) \\
\theta &\sim \pi(\mm{\theta}),
\end{align*}
where $\pi(\cdot)$ is a generic probability density depending on its arguments.  A natural way to perform inference on these models is to use a Metropolis-within-Gibbs scheme that updates all of   $\mm{x}$ and all of $\mm{\theta}$ in separate blocks.  The problem with this type of scheme is that $\mm{x}$ and $\mm{\theta}$ are horribly correlated in the posterior and, therefore, the Gibbs sampler will poorly explore the space.  A non-centred parameterisation attempts to reduce the posterior dependence by replacing $\mm{x}$  with a new variable  $ \mm{u} = \mm{F}(\mm{\theta})^T\mm{x} $.  $\mm{F}(\mm{\theta})$ is traditionally taken to be the Cholesky factor of $\mm{Q}(\mm{\theta})$, however other choices are possible.

The methods in this paper, therefore, give a new way to propose from non-centred parameterisations.  It also suggests that good, computationally efficient non-centred parameterisations can be constructed from  traditional preconditioners!  The results in Theorem \ref{thm:converge_LGCP} can also be interpreted as a basic statement about the difference between centred and non-centred parameterisations.  In fact, it corresponds well with the standard understanding that non-centred parameterisations perform well in a ``low information'' context, while centred parameterisations perform well in a ``high information'' situation \citep{NIPS2010_0835}.

\section{Computing the log-likelihood} \label{sec:determinant}

While sampling from a large Gaussian random variable is all that is required for many MCMC proposals, if the latent field $\mm{x}$ in \eqref{LGCP} depends on some unknown parameters, computing the acceptance ratio will require the computation of a ratio of determinants of $\mm{Q}(\mm{\theta})$, where $\mm{\theta}$ are now some unknown parameters.  Unlike when using Cholesky decompositions to sample from Gaussian random vectors, the Krylov methods considered in this paper do not automatically construct an approximation to the log-determinant. While it is difficult to construct efficient, arbitrarily accurate approximations to the log-determinant, there is a straightforward unbiased Monte Carlo estimate.  \citet{bai1996some} used a variant of the Hutchinson estimator \citep{Hutchinson1990} to construct the estimate $$
\log(\det(Q)) = \mathbb{E}_{\mathcal{V}}(\mathcal{V}^T\log(\mm{Q}) \mathcal{V}) \approx N^{-1} \sum_{i=1}^N \mm{v}_i^T\log(\mm{Q})\mm{v}_i,
$$
where the components of $\mm{v}_i$ are i.i.d. with values equal to $\pm 1$ with equal probability, that is, $\mathcal{V}$ is a vector with i.i.d. Rademacher components.

\citet{Hutchinson1990} showed that the above estimator is unbiased and the choice of random vectors gives the minimum variance estimator amongst all centred, uncorrelated random vectors.  However, the variance, which is $\norm{\log(\mm{Q})}_F^2 - \tr(\log(\mm{Q}))^2$,  can still be unacceptably large in practical situations. We suggest a combination of two novel techniques to help reduce the variance to a more manageable level.  The first method to reduce the variance was introduced by \citet{Aune2012} using a combination of \emph{ad hoc} reasoning and numerical experimentation.  In the following paragraphs, we will re-derive this method rigorously  and show that the resulting variance reduction is due to the structure of $\mm{Q}$.  We will use the insight built in this derivation to show how to build preconditioners that maintain the efficiency of the method of \citet{Aune2012} as the dimension of the problem increases.

\subsection{Understanding the coloured Hutchinson estimator of \citet{Aune2012}}

In order to ease notation, we will use the notation $\mm{B} = \log(\mm{Q})$ in the remainder of this section and, for the sake of simplicity, we will assume for the rest of this section that $\mm{Q}$ is sparse. For a realisation $\mm{v}$ of the vector valued random variable $V$, then, by the symmetry of $\mm{B}$, $$
\mm{v}^T\mm{Bv} = \tr(\mm{B}) + 2\sum_{i<k} v_{i}v_j B_{ij},
$$ where $v_i$ is the $i$th component of $\mm{v}$ and $B_{ij}$ is similarly defined.   It is clear that the off-diagonal elements of $\mm{B}$ are the source of the Monte Carlo error.   The off diagonal elements of $\mm{B}$ are not arbitrary.  \cite{benzi2007decay} proved that the entries decay exponentially in the graph distance $d(i,j)$.    Let the eigenvalues of $\mm{Q}$ be contained in the interval $[\lambda_{\min},\lambda_{\max}]$.  Then a combination of Theorem 3.4 and the discussion in Section 3.7 of \citet{benzi2007decay} show that, for any $1< 2R <  2R^*$, 
\begin{equation} \label{decay_bound}
B_{ij} \leq \frac{2}{1-1/(2R)} \max_{t=\pm(R + 1/(4R))} \left|\log\left(\frac{1}{2} \left((\lambda_{\max}-\lambda_{\min})t + \lambda_{\max}+\lambda_{\min}\right)\right)\right|(2R)^{-d(i,j)},
\end{equation} 
where $R^*$ is the smallest value larger than one for which the bound is undefined.

Given that the off-diagonal elements of $\mm{B}$ pollute the Hutchinson estimator, and given that the off diagonal elements of $\mm{B}$ decay geometrically, it makes sense to decompose $\mathcal{V}$ in order to avoid the large elements.  Formally, the strategy of \citet{Aune2012} to reduce the variance of the Hutchinson estimator is to decompose the random vectors as $\mathcal{V} = \bigoplus_{\mathfrak{c} \in \mathcal{C}} \mathcal{V}^\mathfrak{c}$, where $\mathcal{C}$ is a non-overlapping partition of $\{1,\ldots,n\}$ and $\mathcal{V}^\mathfrak{c}$ is a random vector in which the $i$th component has an independent Rademacher distribution if $i \in \mathfrak{c}$ and is zero otherwise. The coloured Hutchinson estimator is then $$
\log(\det(\mm{Q})) = \mathbb{E}(\mathcal{V}^T\log(\mm{Q}) \mathcal{V}) = \sum_{\mathfrak{c} \in \mathcal{C}} \mathbb{E}((\mathcal{V}^\mathfrak{c})^T\log(\mm{Q}) \mathcal{V}^\mathfrak{c}) 
$$ and its variance is given by \begin{align}
 \operatorname{Var}(\mathcal{V}^T\log(\mm{Q}) \mathcal{V}) &= \sum_{\mathfrak{c} \in \mathcal{C}}  \operatorname{Var}((\mathcal{V}^\mathfrak{c})^T\log(\mm{Q}) \mathcal{V}^\mathfrak{c}) \notag \\
  &= \sum_{i\neq j \in \mathfrak{c}\times \mathfrak{c}} B_{ij}^2 \notag \\
& \leq C^2  \sum_{i\neq j \in \mathfrak{c}\times \mathfrak{c}}  (2R)^{-2d(i,j)}, \label{variance_decay}
 \end{align}
 where $C$ is the constant (in $d(i,j)$) term in \eqref{decay_bound}.  Following a suggestion by \citet{tang2012probing}, \citet{Aune2012} constructed the partition $\mathcal{C}$ by colouring the graph corresponding to the sparsity structure of $\mm{Q}^p$ for some small number $p$.  This ensures that, for all $i,j \in \mathfrak{c}$, $d(i,j) > p$ and shows that the variance of the coloured Hutchinson estimator will be reduced.

\begin{figure}
\centering
\includegraphics[width=0.7\textwidth]{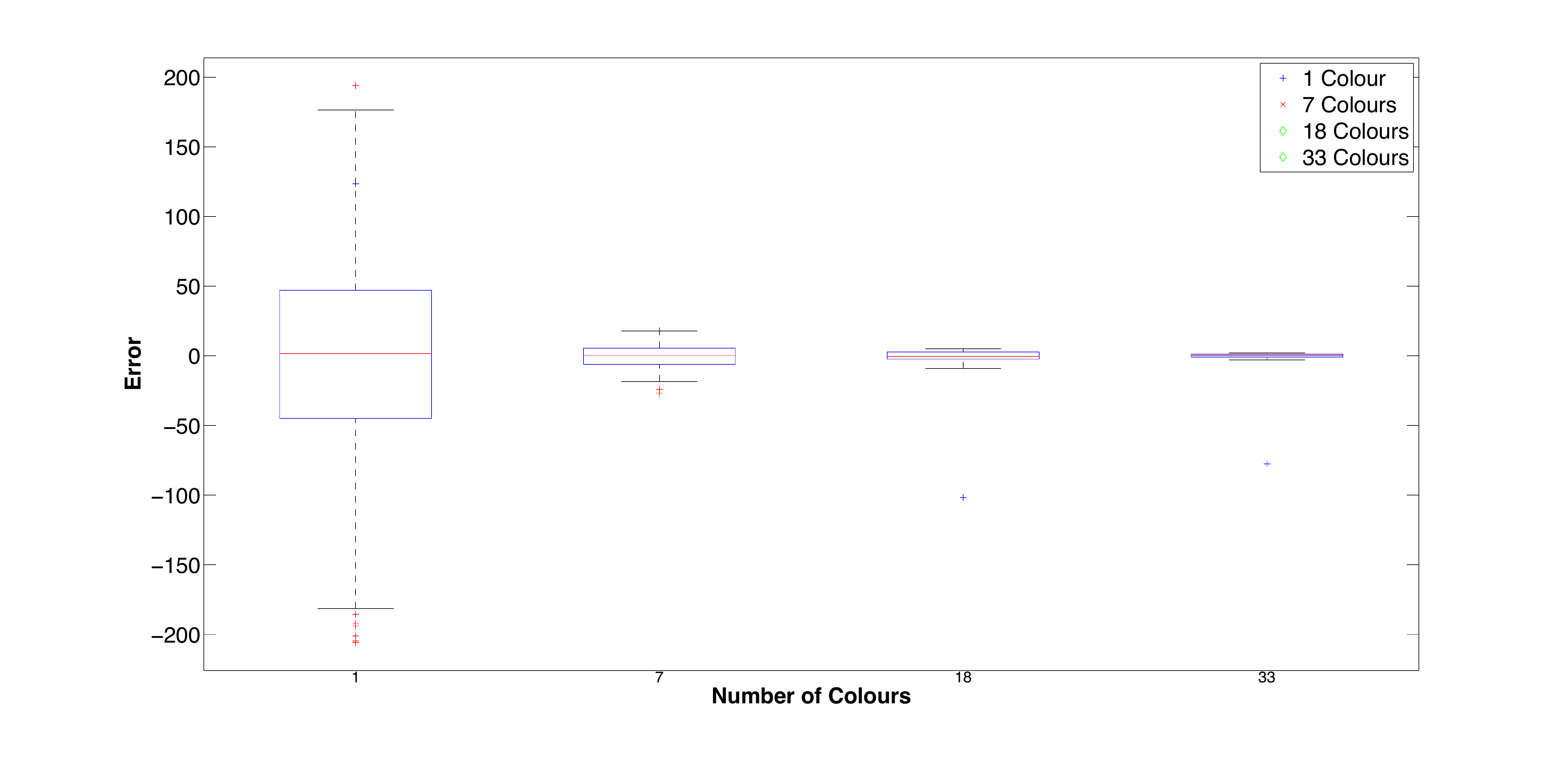}
\caption{This figure shows that a box plot of $\mm{v}^T\log(\mm{Q})\mm{v}$ for the coloured Hutchinson estimator of the matrix considered in Section \ref{sec:general}.  From left to right, there boxes correspond to the uncoloured Hutchinson estimator, the 1-coloured, 2-coloured, and 3-coloured Hutchinson estimators.  It is clear that these are unbiased. \label{fig:variance}}
\end{figure}

The variance reduction and unbiasedness of the coloured Hutchinson estimator is demonstrated in Figure \ref{fig:variance}.  

\subsection{The role of preconditioning}

Unfortunately, the bound \eqref{decay_bound} shows that the there is strong mesh dependence.  Careful analysis shows that $R^* = \frac{1}{2} + \mathcal{O}(\kappa^{-1/2})$, which implies that the distant (in the graph metric) values of $B_{ij}$ become more and more important as the dimension of the problem increases.  Logically, this is not particularly surprising.  If you consider the infill situation that occurs when modelling log-Gaussian Cox processes (the window is fixed, the size of the grid cell decreases), then it is clear that the graph distance does not mirror the physical distance as the dimension of the system increases.  Fortunately, there is a remedy for this.

 Following the theme of the previous section, we note that determinants can also be preconditioned.  In particular, $$
\log \det(\mm{Q}) = \log \det \left(\mm{F}^{-1}\mm{Q} \mm{F}^{-T}\right) - 2 \log \det (\mm{F}).
$$  If  the log determinant of $\mm{F}$ is known by construction, then one only needs to approximate  $\log \det \left(\mm{F}^{-1}\mm{Q} \mm{F}^{-T}\right)$, which should be significantly better behaved than the original problem.   In fact, if $\mm{M} = \mm{FF}^T$ is an optimal preconditioner, than the bound in equation \eqref{decay_bound} suggests that the $\mathcal{R}^*$ is independent of $n$.  This suggests that a good preconditioner will give a similar variance reduction for each colouring regardless of the dimension of the problem.

 In the case of the log-Gaussian Cox process, the proof of Theorem \ref{thm:converge_LGCP} shows that $\mm{Q}_n^{-1} \mm{H}_n$ converges to a trace class operator as $n \rightarrow \infty$ and, therefore, it follows from the theory of Fredholm determinants \citep{bornemann2009numerical} that  $\log \det \left(\mm{F}_n^{-1}\mm{Q}_n \mm{F}_n^{-T}\right)$ stays bounded.  This is important as the computation of the log-acceptance ratio requires the difference of log-determinants and it is computationally unwise to compute the difference of large floating point numbers!

\section{Whither exactness? Balancing the inexactness of  Krylov methods with the inexactness of MCMC}  \label{sec:exactness}

In this paper, we showed that the Krylov methods can be used to compute samples and likelihoods of Gaussian models in a way that can be independent of the dimension of the underlying problem. These methods are necessary in order to perform inference on massive statistical models.  In order to make inference possible on these models, we have sacrificed a degree of exactness and it is reasonable to ask what  effect this has on inference methods.  

Samples computed using Krylov sampling are not Gaussian.  The algorithm works by targeting a fixed sample from a multivariate Gaussian and approximating it within a low-dimensional subspace that is constructed using information from the target sample.  Therefore, rather than computing the linear filter $\mm{Q}^{-1/2}\mm{z}$, we are  instead computing a complicated non-linear function of the Gaussian random vector $\mm{z}$.  The question is then: \emph{does this matter?} From a pragmatic point of view, we  argue that it doesn't.    There are essentially two components to our argument.  The first is an appeal to practicality, where we must admit that by the time that these algorithms are of any use at all, the problems that are being solved are sufficiently difficult that some  sacrifices are needed.  There is also strong evidence that being slightly wrong is not a problem in practical MCMC schemes as the incorrect chain will often follow the ``correct'' chain for a long period of time \citep{nicholls2012coupled}.  The second argument is an appeal to reality. This cardinal rule of numerical computing is that you will never calculate the exact thing that you wish to.  Floating point artefacts pollute even the simplest numerical calculations and in the situations that we have considered in this paper, where we are approximating an infinite dimensional random variable by a high dimensional one, these calculations are anything but simple.  Even an ``exact'' sample computed using a Cholesky factorisation, which is the current gold standard for sampling from large problem, will, in reality, be non-Gaussian due to to the complicated effect of rounding error.  For the sorts of problems considered in this paper, methods based on the Cholesky factorisation will not be exact within floating point tolerance.

A different, but related, difficulty comes from the inexact calculation of the log-determinant.  While this is unbiased, it will lead to a biased estimator of  the acceptance ratio, which is a ratio of determinants. \citet{Girolami2013Playing} showed that it is possible to account for this extra randomness and construct an exact pseudo-marginal MCMC scheme.  However, in the interest of simplicity, we have opted to work with an inexact chain.  Once again, the analysis of \citet{nicholls2012coupled} strongly suggests that inexact MCMC schemes that simply use the estimate directly (or those that  adjust for the variance in the estimator) will lead to methods that are, for all intents and purposes, exact.  They argue that the inexact chain will be coupled with the true chain for, on average, an amount of time proportional to the inverse variance of the estimator.  This means that if the estimator is sufficiently precise, the error cause by the inexactness of the chain will most likely be swamped by the Monte Carlo error.  

The fundamental question, then, becomes not one of accuracy and asymptotic exactness, but rather one of finite sample behaviour.  Ideally, we would like to have a detailed theory that links the accuracy of each step of the MCMC scheme with the finite sample error.  Outside of statistics, this is analogous to the analysis of \citet{dembo1982inexact} for Newton's method for solving systems non-linear equations, in which it was shown that the quadratic convergence of Newton's method can be maintained even when the correction term in computed inexactly.  To the best of our knowledge, the general version of this problem has not been solved, however the recent work of \citet{ketelsen2013hierarchical}, in which detailed work was done to balance error and cost for a class of inverse problems.

\section{Conclusion}  \label{sec:conclusion} 

The methods presented in this paper open up the possibility of general, dimension independent inference methods.  There are a number of further steps required to make this dream a reality.  First and foremost, a great deal of effort must be put into constructing optimal or almost optimal preconditioners that can be factorised.  In this paper we showed that this is possible, however we limited ourselves to (block) circulant   problems.  While the extension to Toeplitz matrices is straightforward,  block Toeplitz matrices pose a significant challenge.  In this case, optimal preconditioners have a banded block Toeplitz form \citep{serra1994preconditioning}. It is possible to sample from the optimal  in $\mathcal{O}( (n\log n)$ operations by combining multigrid methods with a rational approximation to the square root \citep{hale2008computing} and to compute determinants in $\mathcal{O}(n^{3/2}\log(n))$  operations \citep{bini1988efficient}.  These preconditioners are, therefore, much more expensive than in the circulant case.

This highlights a fundamental challenge for this methodology.  At the current time, the gold-standard method for constructing mesh-independent preconditioners for general spatial problems is to use some variant of multigrid, however this procedure cannot be used in our context as it is not possible to sample from a Gaussian vector with a precision operator given by the multigrid operation.   A more promising option may be symmetric sweep Schwartz iterations, which have a strong connection with overlapping block Gibbs samplers for sampling from large Gaussian random vectors.  Another  challenge is to construct preconditioners that have mesh-independent decay properties.  This is needed to ensure that the variance of the preconditioned determinant calculation can be bounded independent of the dimension of the problem.  It is also important to design and study ``algebraic'' preconditioners (such as incomplete Cholesky factorisations) that can be applied to general problems without using some underlying analytical structure of the problem.

The second challenge that must be met for these methods to implement these methods in fast, efficient ways.  \citet{Aune2011} has considered the use of GPUs for sampling from large Gaussian random vectors, while \citet{art192} has used shared memory parallelisation to speed up the INLA software \citep{art451}.  \citet{Paciorek2013Parallelizing} use MPI to distribute the dense linear algebra operations required for inference with general, unstructured covariance matrices.  The methods considered in that paper are limited to small problems, however this type of parallelism is certainly useful for distributing Krylov methods.  

The final challenge is in the design of MCMC samplers that mirror the dimension independence of the Krylov sampler.  There has been some work done in generalising basic samplers like preconditioned Crank-Nicolson \citep{cotter2012mcmc},  preconditioned MALA \citep{beskos2008mcmc}  and Hamiltonian Monte Carlo \citep{beskos2011hybrid} to this case, but this technology has yet to be employed on methods that try to track the local second order properties of the posterior. With these three steps in place, we believe that the methods presented in this paper will have a great impact on inference schemes for massive problems.

\paragraph{Acknowledgements:} The authors would like to thank Erlend Aune, Colin Fox, Al Parker and H\aa{}vard Rue for their feedback and encouragement.

\bibliographystyle{plainnat}
{\bibliography{merged}}

\appendix
\section{Proof of Theorem \ref{thm:converge_LGCP}} \label{appendix:proof}
Without loss of generality, we can take $k=1$.

Noting that $\mathbf{M}_n^{-1}(\mm{Q}_n + \mm{H}_n) =(\mm{Q}_n +\alpha_n\mm{I}_n)^{-1}(\mm{H}_n - \alpha_n \mm{I}_n)$,  if follows that if, $\operatorname{tr}(\mm{Q}_n^{-1}\mm{H}_n)$ is bounded for all $n$, the result follows from   Theorem \ref{thm:convergence} and the results of \citet{axelsson2009equivalent}.  The boundedness of the trace can be shown as follows.
\begin{align*}
\operatorname{tr}\left((\mm{Q}_n +\alpha_n\mm{I}_n)^{-1}(\mm{H}_n - \alpha_n \mm{I}_n)\right) &= \left[(\mm{Q}_n + \alpha_n \mm{I}_n)^{-1}\right]_{1,1}\sum_{i=1}^n ( h^2 e^{x_i} - \alpha_n) \\
& =  \left[(\mm{Q}_n + \alpha_n \mm{I}_n)^{-1}\right]_{1,1}\left( \int_W \exp(x(s))\,ds + O(h^2) - n\alpha_n\right)\\
&\rightarrow C \left(\int_W \exp(x(s))\,ds  - C_\alpha\right) < \infty \, \text{a.s.}
\end{align*}

\end{document}